\documentclass[aps,
		prd,
		twocolumn,
		superscriptaddress,
		shortbibliography,
		nofootinbib,
		floatfix,
		notitlepage
		]{revtex4-2}
\pdfoutput=1

\usepackage[utf8]{inputenc}

\usepackage{amsmath,amssymb,amsfonts}

\usepackage[dvipsnames]{xcolor}

\usepackage{bm}
\usepackage{amsmath}
\usepackage{amssymb}
\usepackage{graphicx}
\usepackage{hyperref}
\usepackage{cleveref} 
\usepackage{slashed}
\usepackage{amsfonts}
\usepackage{amsbsy}
\usepackage{color}
\usepackage{array}
\usepackage{dcolumn}
\usepackage{slashed}
\usepackage{braket}
\usepackage{mathrsfs}
\usepackage{tikz}
\usepackage{verbatim}
\usepackage{tcolorbox}
\usepackage[normalem]{ulem}

\newcommand{\vphi}{\varphi}
\newcommand{\bea}{\begin{eqnarray}}
\newcommand{\eea}{\end{eqnarray}}
\newcommand{\vkap}{\varkappa}
\newcommand{\eps}{\varepsilon}

\newcommand{\nn}{\nonumber}
\newcommand{\tsf}[1]{\textsf{#1}}
\newcommand{\trm}[1]{\textrm{#1}}

\newcommand{\mbf}[1]{\mathbf{#1}}
\newcommand{\Ecr}[1]{E_{\tiny\tsf{qed}}}

\newcommand{\alphaqed}{\alpha_{\tiny\tsf{qed}}}

\newcommand{\figref}[1]{Fig. \ref{#1}}
\newcommand{\figrefa}[1]{Fig. \ref{#1}a}
\newcommand{\figrefb}[1]{Fig. \ref{#1}b}
\newcommand{\figrefc}[1]{Fig. \ref{#1}c}
\newcommand{\figrefs}[2]{Fig. (\ref{#1}) and (\ref{#2})}
\newcommand{\eqnref}[1]{Eq. (\ref{#1})}

\newcommand{\appref}[1]{App. \ref{#1}}

\newcommand{\be}{\begin{equation}}
\newcommand{\ee}{\end{equation}}
\newcommand{\bi}{\begin{itemize}}
\newcommand{\ei}{\end{itemize}}

\newcommand{\ud}{\mathrm{d}}
\newcommand{\LCperp}{{\scriptscriptstyle \perp}}

\begin{document}

\title{
Feasibility of measuring non-analytic QED coupling from pair creation in strong fields
}

\author{B. King}
\email{b.king@plymouth.ac.uk}
\affiliation{Centre for Mathematical Sciences, University of Plymouth, PL4 8AA, UK}

\author{S. Tang}
\email{tangsuo@ouc.edu.cn}
\affiliation{College of Physics and Optoelectronic Engineering, Ocean University of China, Qingdao, Shandong, 266100, China}

\begin{abstract}
In the quasistatic regimes of nonlinear Breit-Wheeler and trident pair creation, the rates can exhibit a non-analytic dependency on the fundamental coupling of quantum electrodynamics (QED), in a form similar to Schwinger vacuum pair creation. To reach this tunneling regime requires satisfying competing requirements: high intensity but low strong-field parameter with sufficient pair creation to be observed. Using a locally monochromatic approach, we identify the parameter regime where tunneling pair-creation could be measured for the first time in experiment. Studying several scenarios of collisions with focussed Gaussian pulses, including a bremsstrahlung and an inverse Compton source for nonlinear Breit-Wheeler and a Gaussian electron beam for nonlinear trident, we find the position of the tunneling parameter regime to be well-defined and robust.
\end{abstract}

\maketitle

\section{Introduction}
Electromagnetic (EM) fields can be converted into electron-positron pairs via a variety of mechanisms. It was predicted by Breit and Wheeler \cite{breit34} that colliding two real photons with sufficient energy should allow their conversion to an electron-positron pair and this was recently observed with quasi-real photons in ultra-peripheral collisions of heavy ions at ATLAS~\cite{ATLAS:2017fur,ATLAS:2019azn} and CMS~\cite{CMS:2018erd}. Alternatively, photons can be converted to electron-positron pairs in the Coulomb fields of nuclei via the Bethe-Heitler process \cite{bethe34}. These are examples of \emph{linear}, leading-order perturbative processes. In contrast, \emph{nonlinear} Breit-Wheeler involves higher numbers of photons, which can occur when a photon scatters in a coherent field where the centre-of-mass energy is too low for the linear, two-photon process, to proceed. Nonlinear Breit-Wheeler was observed in the \emph{multiphoton} regime in the landmark E144 experiment \cite{burke97,bamber99} as the second part of the nonlinear trident process, in which an electron that collides with an EM pulse emits a photon via nonlinear Compton scattering which subsequently is converted to an electron-positron pair. The LUXE experiment \cite{Abramowicz:2021zja,Abramowicz:2021zja} at DESY and the E320 experiment \cite{chen22} at SLAC plan to employ intense laser pulses to observe nonlinear Breit-Wheeler in the all-order regime where arbitrarily high numbers of photons are involved in the creation of a pair. This regime is often referred to as `non-perturbative' because the coupling of the laser photons to the pair is given by the intensity parameter, $\xi$, which can nowadays routinely exceed unity in experiments \cite{danson19} and so including the interaction perturbatively is no longer accurate. 
The laser photon coupling can be related to the elementary QED coupling with $\xi^{2} = 4\pi\alphaqed\lambdabar_{\tiny\tsf{c}}^{2}\lambdabar n_{l}$, where $\alphaqed = e^{2}/4\pi\hbar c$, $e>0$ is the charge on a positron, $n_{l}$ is the number density of laser photons of wavelength $\lambda$ in the background, $\lambdabar_{\tiny\tsf{c}}=\hbar/mc$ the reduced Compton wavelength and $\lambdabar=\lambda/2\pi$ \cite{Abramowicz:2019gvx}. Although $\alphaqed \ll 1$, as the \emph{density} of photons increases, the $\xi \gtrsim O(1)$ regime is accessible. We note that the $\hbar$ dependency in $\alpha_{\tsf{qed}}$ and $n_{l}$ cancels with the $\hbar$ dependency of $\lambdabar_{\tiny\tsf{c}}^{2}$; $\xi$ is of classical origin. This `non-perturbativity at small coupling' has analogies with other fields, such as gluon saturation in inelastic scattering \cite{Gribov:1983ivg,STAR:2021fgw}.

The pair creation process referred to as Schwinger pair creation \cite{Schwinger:1951nm}, which was already conceived by Sauter \cite{Sauter:1931zz} and calculated by Heisenberg and Euler \cite{Heisenberg:1936nmg}, is often contrasted to the Breit-Wheeler process since it traditionally considers pairs produced directly from a constant or slowly-varying electric field. However, more realistic scenarios for observing this process in experiment involve `assisting' Schwinger \cite{Schutzhold:2008pz,Orthaber:2011cm,Hebenstreit:2011wk,Kohlfurst:2012rb,Schneider:2014mla,Torgrimsson:2016ant,Torgrimsson:2017cyb,Aleksandrov:2018uqb,XIE2017225,OLUGH2020135259,PRD036015,Ilderton:2021zej} by adding a high frequency but low-intensity field. Therefore, we might expect that in some parameter regime there is similarity to nonlinear Breit-Wheeler which considers a slowly-varying background (albeit one that can be well-approximated as a plane wave) colliding with a high energy (frequency) photon. (Recently scenarios have also been considered where nonlinear Breit-Wheeler is `assisted' by a combination of low and high frequency laser pulse \cite{Akal:2014eua,Brass:2019pzr,Folkerts:2023gpx,Mahlin:2023aui}.) Schwinger pair-creation is often highlighted as special, because of the non-analytic dependency on the QED coupling, $\alphaqed$, in a constant electric field. This is a second type of non-perturbativity; unlike the small-coupling case, which is an entirely classical effect, the non-analytic dependence on $\alphaqed$ is a fundamentally quantum non-perturbativity. Therefore it is of interest to observe this type of non-perturbativity as an aim separate to the classical `small-coupling' non-perturbativity when $\xi \gtrsim O(1)$. 

It is known that the nonlinear Breit-Wheeler process, just like the Schwinger process, can exhibit a non-analytic dependency on $\alphaqed$ in the `quasistatic' or `locally constant' regime. In nonlinear Breit-Wheeler, the key parameter is $\chi$, which for a charge in a plane wave background is exactly $\chi = E_{\tiny\tsf{r.f}}/E_{\tiny\tsf{qed}}$ i.e. the ratio of the field strength in the charge's rest frame, $E_{\tiny\tsf{r.f}}$, to  the QED field strength scale (`Schwinger limit'), $E_{\tiny\tsf{qed}} = m^{2}c^{3}/e\hbar$. Since $\chi \propto \hbar$, we see it is a quantum parameter and therefore cannot be acquired from $\xi$. The QED coupling then appears in the nonlinear Breit-Wheeler rate, in the square root of the denominator of an exponent, just as in the Schwinger process. We will refer to this as the \emph{tunneling regime} for simplicity (even though the potential is a plane wave and not static). It was suggested in \cite{Reiss:1971wf,Hartin:2018sha} to measure nonlinear Breit-Wheeler in the tunneling regime by combining a bremsstrahlung photon source with a monochromatic laser and vary its intensity. However, to reach the tunneling regime with a fixed photon energy, opposing limits must be fulfilled: $\xi \gg 1$ and $\chi \ll 1$ whilst also producing sufficient pairs to be measurable. Furthermore, the tunneling regime is an \emph{asymptotic} limit and so it is unclear before calculation, what the magnitude of parameters must be for the process to be well-described by this regime.

In the current paper, we outline the limited but accessible parameter space in which experiments could observe the non-analytic dependency on $\alphaqed$ in nonlinear pair-creation. This region crucially depends on photon energy as well as EM field intensity, and is reachable at high intensity laser facilities \cite{apollon16,corels17,eliBeamlines17} as well as at laser-particle experiments LUXE \cite{Abramowicz:2021zja, Abramowicz:2023sxm} and E320 \cite{chen22}. We cannot use the standard locally constant field approximation (LCFA) to calculate the rate of strong-field QED processes, because the tunneling regime is a limit of this approximation and so a more accurate calculational framework that does not rely upon locally constant rates is required. For this reason, we employ in our analysis the locally monochromatic approximation \cite{Heinzl:2020ynb} (LMA), which has been benchmarked against exact plane wave calculations \cite{Blackburn:2021cuq,Tang:2021qht} and incorporated into the open source Ptarmigan simulation code \cite{Blackburn:2023mlo,ptarmigan}. We employ Ptarmigan's rate generator for pair creation, which allows efficient calculation of the LMA at high intensities. Our aim is to assess the feasibility of measuring pair creation in the tunneling regime, and to this end we consider scenarios that include probe beam distributions and focussed laser backgrounds.

We organise the investigation by introducing increasingly detailed modelling of potential experiments in each successive section. This begins with the general situation in Sec.~\ref{Sec2}, with monoenergetic photons colliding with a plane-wave background. In Sec.~\ref{Sec3}, the monoenergetic photons are replaced with a photon source from: i) an amorphous bremsstrahlung target; ii) inverse Compton-scattering, and the plane-wave background is replaced with a Gaussian focussed laser pulse. In Sec.~\ref{Sec4} we investigate a scenario for measuring the non-analytic coupling using an electron probe in the two-step nonlinear trident process. In Sec.~\ref{Sec5} we conclude the feasibility study of measuring Schwinger-like pair-creation using intense lasers.
(Unless otherwise stated, we set $\hbar=c=1$ throughout.)

\section{Monoenergetic photons, plane wave background}~\label{Sec2}
Here we consider monoenergetic photons colliding with a plane-wave background which demonstrates the essence of the problem. The tunneling rate, $\tsf{P}_{\tiny\tsf{tun}}$, for pair-creation via nonlinear Breit-Wheeler is \cite{ritus85}:
\bea
\frac{d\tsf{P}_{\tiny\tsf{asy}}}{d\vphi} = \frac{\alphaqed}{\eta}\frac{3}{16} \sqrt{\frac{3}{2}}\,\chi \,\exp\left[-\frac{8}{3\chi}\right] \label{eqn:asy1}
\eea
where $\vphi = \vkap \cdot x$ is the plane-wave phase and the energy parameter of the photon with momentum $\ell$ is $\eta=\vkap \cdot \ell / m^{2}$. We compare this to the rate of pair creation via the Schwinger mechanism \cite{narozhny04,fedotov09}:
\bea
\frac{d\tsf{P}_{\tiny\tsf{Sch.}}}{d^{4}x} = \frac{\alphaqed}{4\pi^2} \mathfrak{a}\mathfrak{b}\,\trm{coth}\left(\frac{\pi \mathfrak{b}}{\mathfrak{a}}\right)\,\exp\left[-\dfrac{\pi E_{\tiny\tsf{qed}}}{\mathfrak{a}}\right], \label{eqn:Schw1}
\eea
where the secular field invariants are:
\bea
\mathfrak{a},\mathfrak{b} = \left[\sqrt{\mathcal{S}^{2}+\mathcal{P}^2}\pm \mathcal{S}\right]^{1/2},
\eea
and $\mathcal{S} = -F_{\mu\nu}F^{\mu\nu}/4$ with $\mathcal{P} = -\widetilde{F}_{\mu\nu}F^{\mu\nu}/4$ are the standard EM invariants where $F$ is the field tensor and $\widetilde{F}$ is its dual. We note the common non-analytic dependncy on the fundamental QED coupling of these rates since $1/\chi \propto 1/\sqrt{\alphaqed}$ in \eqnref{eqn:asy1} and $E_{\tiny\tsf{qed}} \propto 1/\sqrt{\alphaqed}$ in \eqnref{eqn:Schw1}. (Breit-Wheeler is sometimes distinguished from Schwinger with the argument that $\mathcal{S}$ and $\mathcal{P}$ are zero in a plane wave background and hence the contribution from the Schwinger process is zero. In \appref{app:finvariants} we note that when, in the definition of $F$, the photon is taken into account as well as the plane-wave background, $\mathcal{S}$ and $\mathcal{P}$ are clearly non-zero and in fact proportional to $\chi$.)

The tunneling rate in \eqnref{eqn:asy1} is the $\chi \to 0$ asymptotic limit of that in a locally constant EM background. The locally constant field approximation (LCFA) is expected to be a good approximation when $\xi \gg 1$, but for pair-creation also requires that the photon energy parameter, $\eta$, is not too high \cite{Blackburn:2021cuq} (otherwise pair-creation can proceed via the multiphoton process, which is not captured by the LCFA). Therefore for the tunneling rate in \eqnref{eqn:asy1} to be accurate, we have two limits: i) $\xi \to \infty$ for the LCFA to be accurate and ii) $\chi \to 0$ to be in the tunneling limit. However, since $\chi = \eta \xi$, these two limits are not independent. It has been shown that in the high-$\chi$ limit, the $\xi \to \infty$ and $\chi \to \infty$ limits do not commute \cite{Ilderton:2019kqp,Podszus:2018hnz}; here we find ourselves in a similar situation but a different limit. This is essentially due to $\chi$ being a compound parameter, depending on the product of particle and field parameters, which are usually considered independently in experiment.

The situation is further complicated by vague language such as `good approximation' and inexact limits like `$\xi \gg 1$, $\chi \ll 1$'. In this article we will take a practical approach: we will calculate using the more widely applicable and accurate locally monochromatic approximation (LMA) and compare with the asymptotic tunneling formula in \eqnref{eqn:asy1}. If the scaling of the pair creation rate with $\chi$ and hence with $\alphaqed$ is described by the tunneling formula to within some error (which we nominally take to be $10\%$), then we conclude in that parameter regime, that pair-creation exhibits the predicted non-analytic dependency on $\alphaqed$. Then the question remains: `how high must $\xi$ be?' and `how small must $\chi$ be?' to see this dependency in experiment, i.e. whilst producing sufficient pairs as to be measurable.
\newline

\begin{figure}[h!!]
\includegraphics[width=0.7\linewidth]{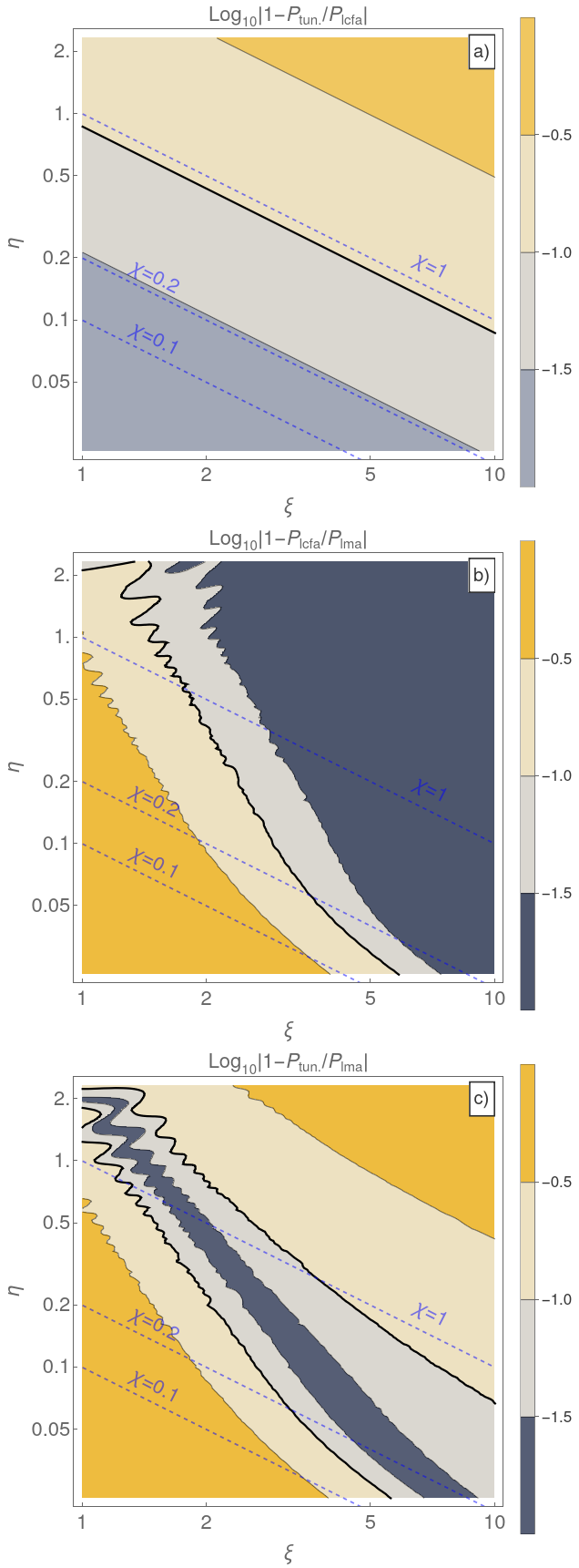}
\caption{Relative probabilities for nonlinear Breit-Wheeler pair-creation in a four-cycle plane-wave pulse. The thick contour lines denote a relative difference of $10\%$.} \label{fig:monoPlot}
\end{figure}

To investigate this, we consider monoenergetic photons with energy parameter $\eta$ colliding with a plane wave with scaled potential $a = eA$:
\bea
a(\vphi) = m \xi \eps \sin^{2}\left(\frac{\vphi}{2N}\right)\,\cos\vphi \label{eqn:pw1}
\eea
for $0<\vphi<2\pi N$ where $N$ is the number of cycles and $a(\vphi)=0$ otherwise, and $\eps$ is a linear polarisation vector. We can write the ratio of the tunneling and LMA probability as:
\bea
\frac{\tsf{P}_{\tiny\tsf{tun.}}}{\tsf{P}_{\tiny\tsf{lma}}} = \frac{\tsf{P}_{\tiny\tsf{tun.}}}{\tsf{P}_{\tiny\tsf{lcfa}}}\times \frac{\tsf{P}_{\tiny\tsf{lcfa}}}{\tsf{P}_{\tiny\tsf{lma}}}.
\eea
The first ratio on the right-hand side is a measure of how close the LCFA is to the tunneling regime, and the second ratio is a measure of how close the LCFA is to the LMA and the true probability. We can use this relation to understand when the tunneling regime is arrived at, by plotting in \figref{fig:monoPlot} how close each of these quantities is to unity.
(The expressions for $\tsf{P}_{\tiny\tsf{lcfa}}$ and $\tsf{P}_{\tiny\tsf{lma}}$ can be found in App.~\ref{APP_LCFA_LMA}.)

In \figrefa{fig:monoPlot}, the relative difference of the tunneling formula to the LCFA is plotted, showing when the latter is in the tunneling regime. We note that the accuracy increases in a direction parallel to falling $\chi$, as expected.

In \figrefb{fig:monoPlot}, the relative difference of the LMA to the LCFA is plotted. Broadly speaking, there is better agreement when $\xi \gg 1$, but there is a slight tilt with higher energies being more accurate for pair creation.

The main result is in \figrefc{fig:monoPlot}. Here we see that if $\xi$ is too small, the tunneling formula is not accurate because the LCFA is not accurate (as reflected by \figrefb{fig:monoPlot}). Likewise, if the energy is too large, the $\chi$ value is too large and so the tunneling limit is not reached (as reflected by \figrefa{fig:monoPlot}). Therefore, there is an optimum region in the middle of the plot. For $\eta=0.1$, we see the region is optimal for the relative difference smaller than $10\%$ for $0.3\lesssim\chi\lesssim 0.8$, i.e. $3\lesssim \xi \lesssim 8$, which can correspond to a measurable number of pairs. (In the optimal region, the relative difference between tunneling and LMA probabilities changes sign ).

\begin{figure}[h!!]
\includegraphics[width=0.8\linewidth]{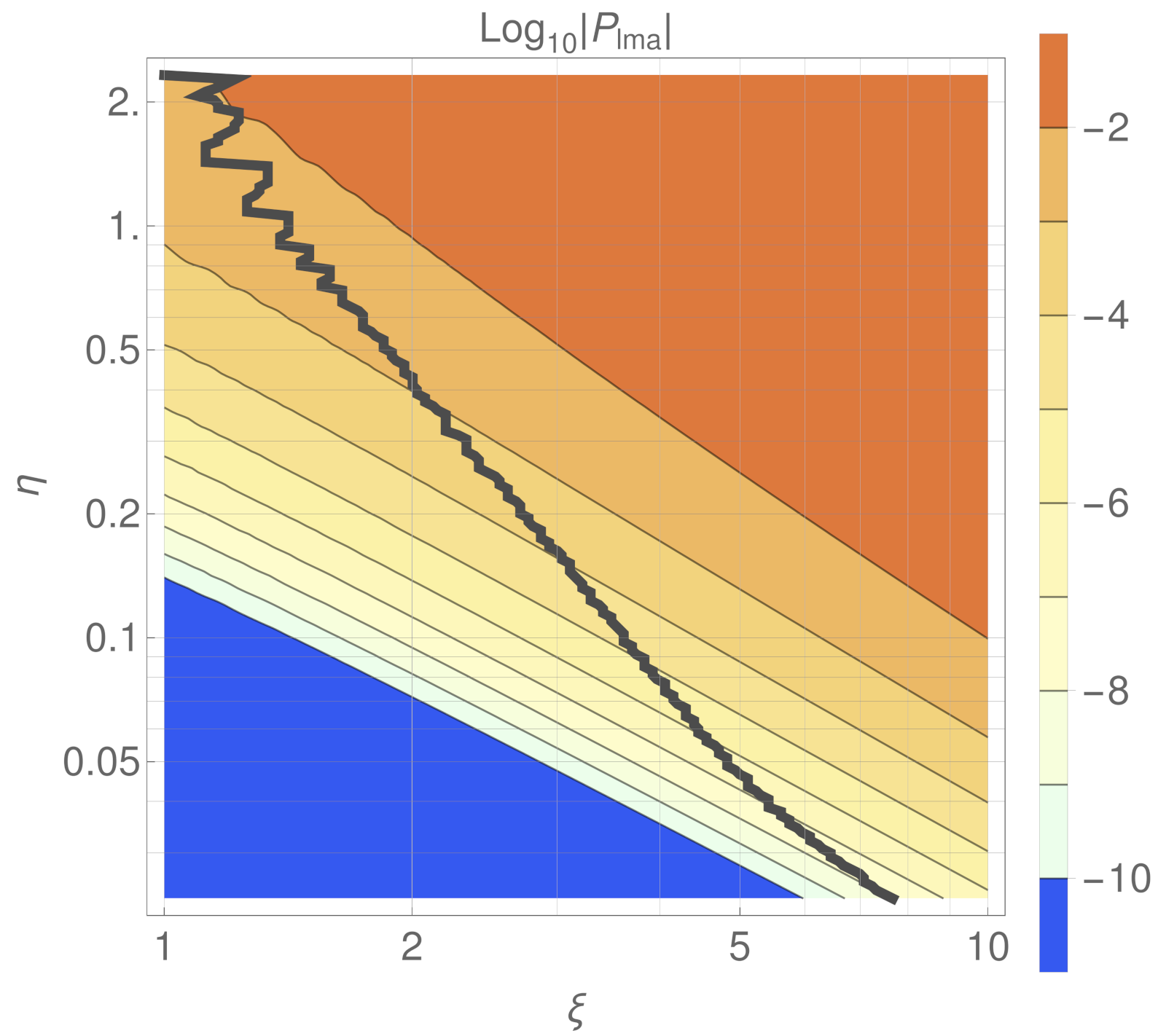}
\caption{Probability for pair creation for the monoenergetic photon colliding with a $16$-cycle plane-wave. The dark grey line is the `tunneling' region from \figrefc{fig:monoPlot}.} \label{fig:monoPlotProb}
\end{figure}

For a given photon energy, \figrefc{fig:monoPlot} shows the optimal EM background intensity to be in the tunneling regime. However it is only feasible to measure in this regime if sufficient pairs are generated. In \figref{fig:monoPlotProb}, the `tunneling regime' curve is plotted on top of the probability for pair creation. The optimal parameters for detecting tunneling pair creation can then be seen to be in the `transition regime' of intensity $\xi \gtrsim O(1)$ at higher photon energies $\eta \gtrsim O(0.1)$. It is clear that moving to higher intesities and lower energies, whilst being in the optimal tunneling regime, will lead to too few pairs being created for experiments, unless they involve very large number of initial high energy photons.

\section{photon source}~\label{Sec3}
In the cases of photons from bremsstrahlung and inverse Compton sources colliding with a plane wave, there is very little modification of the conclusions in  \figref{fig:monoPlot}. In the bremsstrahlung case, although the photon spectrum is broad, since pair creation in the tunneling regime depends strongly on the photon energy, the main contribution is just from the highest photon energies. The only difference is in the harmonic structure, which can be seen in the high-$\eta$, small-$\xi$ oscillations in the probability in \figrefs{fig:monoPlot}{fig:monoPlotProb}. For the inverse Compton source, the bandwidth is narrow and quasi mono-energetic. Therefore we move straight to the focussed background case.

\subsection{Bremsstrahlung photons}
\begin{figure}[h!!]
\includegraphics[width=0.8\linewidth]{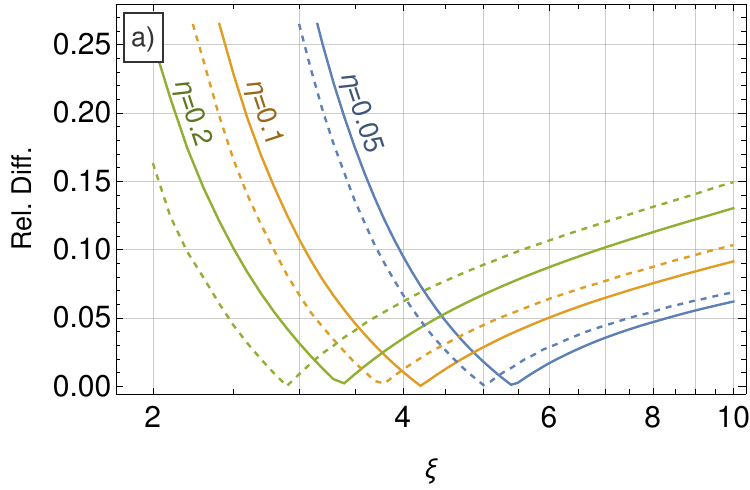}\\
\includegraphics[width=0.8\linewidth]{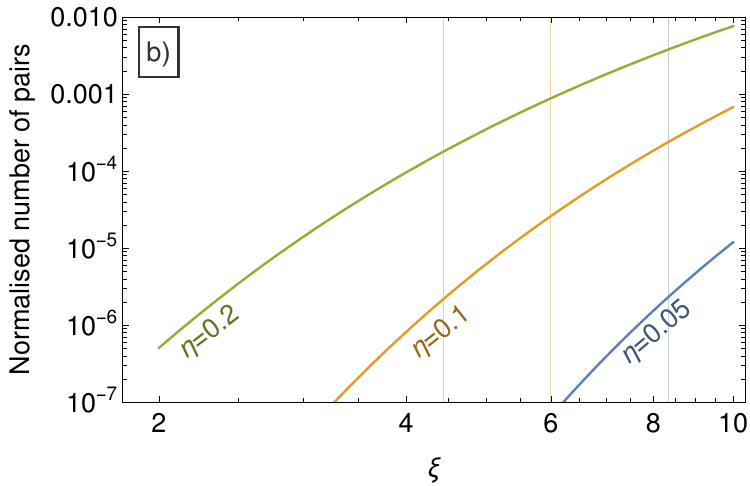}
\caption{Fig a): The relative difference of the tunneling and the LMA calculations. For lines of the same colour: the dashed lines are for a plane wave and the solid lines are for the focussed Gaussian beam in the infinite Rayleigh length approximation. Fig b): The normalised number of pairs. The gridlines with the same colour as a curve indicate the upper bound of being within $5\%$ of the tunneling formula.}
\label{fig:focussed1}
\end{figure}

In this section, we calculate the probability for pair creation, $\tsf{P}^{\tiny\tsf{brems}}$, in the collision of thin-target bremsstrahlung with a focussed laser pulse. To do this, we integrate the plane-wave probability, $\tsf{P}^{\tiny\tsf{pw}}$, for nonlinear Breit-Wheeler (calculated with the LMA or the tunneling formula) over the impact parameter of the photons in the bremsstrahlung:
\bea
\tsf{P}^{\tiny\tsf{brems}} = \int_{0}^{1} ds\, \int d^{2}\mbf{x}^{\perp}~\rho_{b}\left(s, \mbf{x}^{\perp}\right)~\tsf{P}^{\tiny\tsf{pw}}\left[s,\xi\left(\mbf{x}^{\perp}\right)\right] \label{eqn:Pwidth}
\eea
where $\rho_{b}(s,\mbf{x}^{\perp})$ is a number density of bremsstrahlung photons, $s=\omega_{\gamma}/p^{0}$ is the ratio of photon to initial electron beam energy (assumed quasi monoenergetic). The equation in (\ref{eqn:Pwidth})
 uses a semiclassical approximation assuming that $\gamma \gg \xi^{2}$ where $\gamma$ is the created particles' Lorentz gamma factor (see e.g. \cite{DiPiazza:2016maj}). To calculate $\tsf{P}$ we take the example of a paraxial Gaussian beam in the infinite Rayleigh length limit:
\bea
\xi\left(\mbf{x}^{\perp}\right) = \mbox{e}^{-\frac{|\mbf{x}^{\perp}|^{2}}{w_{0}^{2}}}\,\xi,\label{Eq_Gaussian_beam}
\eea
where $\xi$ is the same as in the definition of the plane wave \eqnref{eqn:pw1}. For the bremsstrahlung, the laser interaction point is typically far enough from the bremsstrahlung source, that over width of the focus of the laser, the bremsstrahlung density can be considered to be independent of transverse co-ordinate. Then we approximate $\rho_{b}\left(s, \mbf{x}^{\perp}\right) \approx \rho_{b}\left(s\right)$ where \cite{Tsai:1974}:
\bea
\rho_{b}(s) = \frac{1}{s}\left[\frac{4}{3}-\frac{4}{3}s+s^{2}\right].
\eea
The results of the numerical calculation can be see in \figrefa{fig:focussed1}, where the vertical axis is the relative difference $|\tsf{P}^{\tiny\tsf{brems}}_{\tiny\tsf{lma}}/\tsf{P}^{\tiny\tsf{brems}}_{\tiny\tsf{tun}}-1|$ where $\tsf{P}^{\tiny\tsf{brems}}_{\tiny\tsf{lma}}$ ($\tsf{P}^{\tiny\tsf{brems}}_{\tiny\tsf{tun}}$) is the evaluation of \eqnref{eqn:Pwidth} with the LMA (tunneling formula) for $\tsf{P}^{\tiny\tsf{pw}}$. For the photon energies considered, which are typical for possible laser-particle experiments, the optimal region for measuring tunneling pair creation is approximately in the range $\xi\in[3,7]$ (for a 20 degree collision with an optical $800\,\trm{nm}$ laser pulse, $\eta\in\{0.05,0.1,0.2\}$ corresponds to photon energies $\{4.3, 8.7, 17.3\} \trm{GeV}$ respectively).

However, we also see in \figrefb{fig:focussed1} that if lower photon energies are considered, for a detectable number of pairs to be generated, the intensities must be increased to the point that $\chi \not \ll 1$ and the process cannot be considered tunneling anymore. This can be seen by noting what the maximum number of pairs per photon is, for the tunneling formula to be accurate to within $5\%$ as indicated by the vertical gridlines in \figrefb{fig:focussed1}. Although the optimal intensity (and spread of intensities) of the `tunneling' regime increases with decreasing photon energy, the number of pairs produced in this optimal regime falls significantly. For comparison, in the LUXE experiment, with $\eta=0.192$, the number of bremsstrahlung photons that collide with the laser focus is of the order of $10^5$ \cite{Abramowicz:2021zja}.

\subsection{Inverse Compton source}
By combining an electron beam with low energy spread with a weak ($\xi \ll 1$) laser pulse of broad focus, a narrowband source of high-energy photons can be produced via inverse Compton scattering \cite{Aoki:2004sr,weller2009research,Albert10,Rykovanov:2014ira,Seipt:2014yga}. The parameters we choose in this section for the laser pulse and electron beam are motivated by the LUXE experiment \cite{Abramowicz:2023sxm}, for which it is planned to frequency-tripling the optical laser and generate a narrowband source of Compton-scattered photons with energies as high as $\sim 9\,\trm{GeV}$.

The inverse Compton source (ICS) can be modelled using the collision between a beam of high-energy electrons (with momentum $p^{\mu}$) and a weak laser pulse \cite{PRAB034401}, giving the double differential spectrum~\cite{BenPRA2020}
\begin{align}
\frac{\ud^{2}\tsf{P}_{\gamma}}{\ud s\ud \theta} 
=&\frac{\alphaqed }{(2\pi)^2} \frac{m^2 /\omega^{2}_{l}\sin\theta}{(1+\cos\theta)^2}\frac{s}{1-s}\int^{\pi}_{-\pi} \ud \psi\iint \ud\vphi_{1} \ud\vphi_{2}\nonumber\\
 &e^{ i\int^{\vphi_{1}}_{\vphi_{2}}\ud \phi\frac{\ell\cdot \pi_p/m^{2}}{ \eta_{e} (1-s))}}\left[\frac{(a(\vphi_{1}) -a(\vphi_{2}))^2}{2m^2}h_{s}-1\right]\,,
\end{align}
characterised by the electron energy parameter $\eta_{e} = \vkap\cdot p/m^{2}$ and the intensity $\xi$ of the laser pulse,
where $\ell^{\mu}=\omega_{\gamma}(1,\sin\theta\cos\psi,\sin\theta\sin\psi,\cos\theta)$ is the momentum of the scattered photon,
$\theta$ is the angular spread of the scattered photon along the anti-direction of the laser pulse, $\vkap^{\mu} = \omega_{l}(1,0,0,-1)$,
$s = \vkap\cdot \ell/\vkap\cdot p\approx \omega_{\gamma}/p^{0}$ for nearly head-on collisions,
$\omega_{l}$ is the laser carrier frequency, and $h_{s} = [1+(1-s)^{2}]/(2-2s)$, $\pi_{p}(\phi)=p^{\mu}+a^{\mu}- (p\cdot a/\vkap\cdot p + a^{2}/2\vkap\cdot p)~\vkap^{\mu}$ is the instantaneous momentum of the electron.

\begin{figure}[t!!!]
 \center{\includegraphics[width=0.45\textwidth]{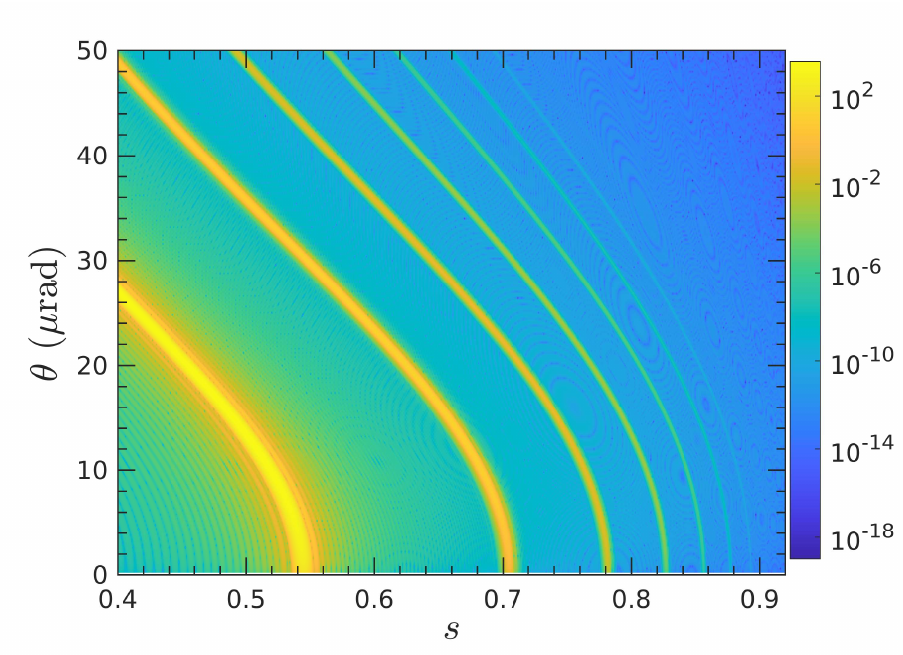}}
         \caption{Double differential spectrum $\ud^{2} \tsf{P}_{\gamma}/(\ud s \ud \theta)$ of the inverse Compton source photons, produced when a $16.8~\trm{GeV}$ electron collides head-on with a laser pulse with sine-squared envelope, carrier frequency $\omega_{l}=4.65~\trm{eV}$ and intensity $\xi=0.1$.  }
\label{Fig_ICS_source}
\end{figure}

In Fig.~\ref{Fig_ICS_source}, we plot the photon distribution of the ICS scattered from a laser pulse with the sine-squared profile used in \eqnref{eqn:pw1}. (Since the weak laser is broadly focussed, it can be approximated during the interaction with the electron beam to be plane-wave in form.) The ICS laser pulse duration is chosen to be  $\trm{FWHM}=25~\trm{fs}$, with carrier frequency (third harmonic) $\omega_{l}=4.65~\trm{eV}$ and intensity $\xi=0.1$, which collides head-on with a $16.8~\trm{GeV}$ electron, corresponding to the energy parameter $\eta_{e}=0.6$. (In the LUXE experiment~\cite{Abramowicz:2021zja}, the electron beam with the energy $16.5~\trm{GeV}$ would be applied.) 
As shown, the ICS photons are quasi-monoenergetic in a narrow angular spread $\theta<10~\mu\trm{rad}$, around $s= 2\eta_{e}/(2\eta_{e} + 1)$ corresponding to the photon energy about $\omega_{\gamma}=9.1~\trm{GeV}$, and the height of the harmonic peaks, around $s= 2n\eta_{e}/(2n\eta_{e} + 1)$ with the integer $n\geq 2$ for the multi-photon scattering, is negligible because of the weak laser intensity.

This ICS photons can be employed to measure tunnelling pair creation by colliding with an intense laser pulse downstream the photon beam.
The creation yield can be written in form as
\begin{align}
\tsf{P}^{\tiny\tsf{ics}}
       &= \iint\ud s\ud \theta~\frac{\ud^{2} \tsf{P}_{\gamma}}{\ud s\ud \theta}~\tsf{P}^{\tiny\tsf{pw}}\left[\omega_{\gamma}(s), \xi(\theta)\right]\,,
\end{align}
where $\tsf{P}^{\tiny\tsf{pw}}$ is the yield of the positrons created by the photon with the energy $\omega_{\gamma}$ and the scattering angle $\theta$.
Again, we consider the infinite Rayleigh length approximation of a Gaussian paraxial beam Eq.~(\ref{Eq_Gaussian_beam}) with the impact parameter $|\mbf{x}^{\LCperp}|=d \tan\theta$ varying with the scattering angle, and also the intensity $\xi(\theta)= \xi\exp(- d^{2}\tan^{2}\theta /w^{2}_{0})$, where $d\sim O(1)\trm{m}$ is the distance between the ICS and the interaction point with the target Gaussian beam.

\begin{figure}[h!!]
\center{\includegraphics[width=0.45\textwidth]{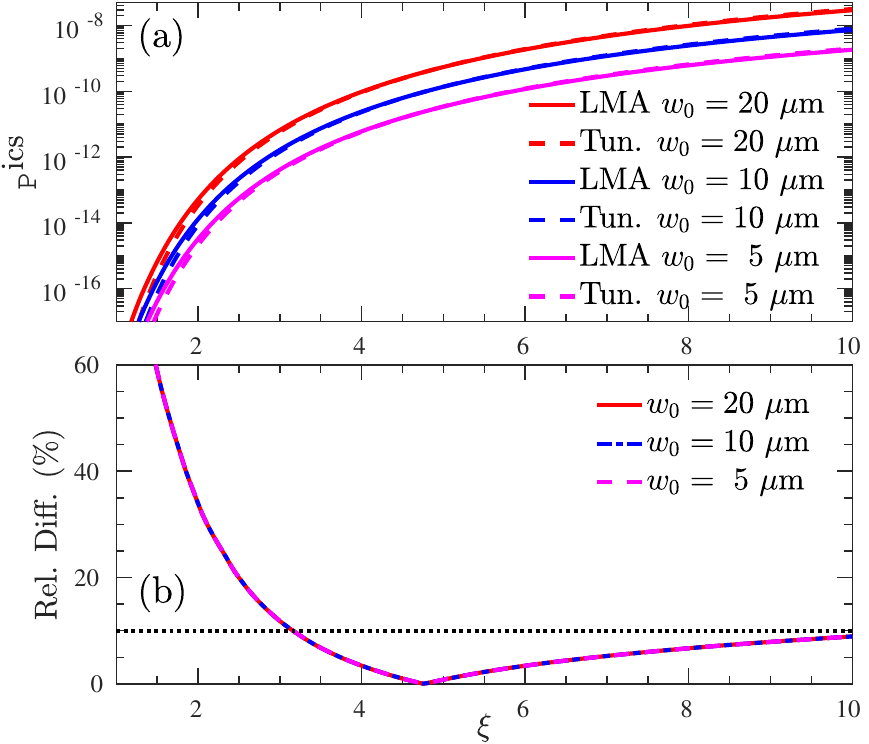}}
         \caption{(a) Positron yield created by the focused pulse with the waist $w=20~\mu\trm{m},~10~\mu\trm{m},~5~\mu\trm{m}$ calculated by the LMA and tunnelling results. (b) Relative difference between the positron yield calculated with each method: $|\tsf{P}^{\tiny\tsf{ics}}_{\tiny\tsf{lma}}-\tsf{P}^{\tiny\tsf{ics}}_{\tiny\tsf{tun}}|/\tsf{P}^{\tiny\tsf{ics}}_{\tiny\tsf{lma}}$.
         The black dotted lines denote the relative difference at $10\%$.
         The Gaussian pulse has the longitudinal profile (\ref{eqn:pw1}) with $16$ cycles.}
\label{Fig_ICS_source_Gausss}
\end{figure}
Fig.~\ref{Fig_ICS_source_Gausss} shows the calculation of the positron yield created by the ICS in Fig.~\ref{Fig_ICS_source} colliding with a Gaussian pulse with the carrier frequency $\omega_{l}=1.55~\trm{eV}$ and $d=7.5~\trm{m}$ downstream, in which the probability of pair creation is calculated with the tunnelling rate [Eq.~(\ref{eqn:asy1})] and the LMA result [Eq.~(\ref{Eq_LMA_NBW}) using the formula in App.~\ref{APP_LCFA_LMA}], respectively. As shown, the positron yield from the LMA calculation increases rapidly with the increase of the laser intensity in the same trend as that calculated with the tunnelling rate, and the relative difference suggests the optimal measurement of the non-analytic dependency pair creation in the intermediate intensity region around $\xi=4.5$, similar to the bremsstrahlung case. We find that the transverse waist $w_{0}$ of the Gaussian pulse only affects the yield of the positron, but not the optimal intensity regime.

\section{Electron beam}~\label{Sec4}
The tunnelling regime of pair creation can also be accessed via the two-step nonlinear trident process~\cite{PRD096004}, in which an electron scatters in a laser pulse to produce a nonlinear Compton scatters which is converted to a pair in the same pulse. The nonlinear trident process thereby combines in the same laser pulse, the photon generation and pair production steps that were separated in the previous section in the bremsstrahlung and ICS set-ups. Furthermore, the two-step process is expected to dominate the total trident rate when $\xi \omega_{l}\tau \gg 1$ \cite{King:2013osa,Mackenroth18} which is generally easily fulfilled  when intense lasers are employed to search for strong-field QED effects.

The positron yield in the two-step trident process can be simply written as
\begin{align}
\tsf{P}^{\tiny\tsf{tri}} &= \int_{0}^{1} \ud s \int^{\vphi_{f}}_{\vphi_{i}} \ud \vphi \frac{\ud^{2}\tsf{P}_{\gamma}}{\ud s \ud \vphi} \int^{\vphi_{f}}_{\vphi} \ud\phi \frac{\ud\tsf{P}^{\tiny\tsf{pw}}}{\ud \phi}\,,
\end{align}
where $\tsf{P}_{\gamma}$ is the probability of the nonlinear Compton scattering calculated also with the LMA result:
\begin{align}
\frac{\ud^{2}\tsf{P}_{\gamma}}{\ud s \ud \vphi}= &\frac{\alphaqed }{\eta_{e}} \sum_{n=\lceil n_{c} \rceil}^{+\infty} \int^{\pi}_{-\pi} \frac{\ud \psi}{2\pi}\nonumber\\
&\left[\xi^{2}(\vphi)\left(\Lambda^{2}_{1,n} - \Lambda_{0,n}\Lambda_{2,n}\right)h_s - \Lambda^{2}_{0,n}\right],
\end{align}
with $\xi(\vphi)=\xi_{0}f(\vphi)$, $\lceil n_{c} \rceil$ denotes the lowest integer greater than or equal to \mbox{$n_{c}=s[1+\xi^{2}(\vphi)/2]/[2\eta_{e}(1-s)]$},
$\Lambda_{j,n}(u,v)$ are the generalised Bessel functions defined in App.~\ref{APP_LCFA_LMA} with $j=0,1,2$ and the arguments given as
\mbox{$u= [r_{c,n} s\xi(\vphi)\cos \psi]/[\eta_{e} (1-s)]$}, $v= s\xi^{2}(\vphi)/[8\eta_{e} (1-s)]$, and $r_{c,n}   = \sqrt{2n\eta_{e}(1-s)/s - 1 - \xi^{2}(\vphi)/2}$.
$\ud\tsf{P}^{\tiny\tsf{pw}}/\ud \phi$ is the rate for the pair creation calculated with the tunneling result and the LMA result.

\begin{figure}[t!!!!!!]
 \center{\includegraphics[width=0.45\textwidth]{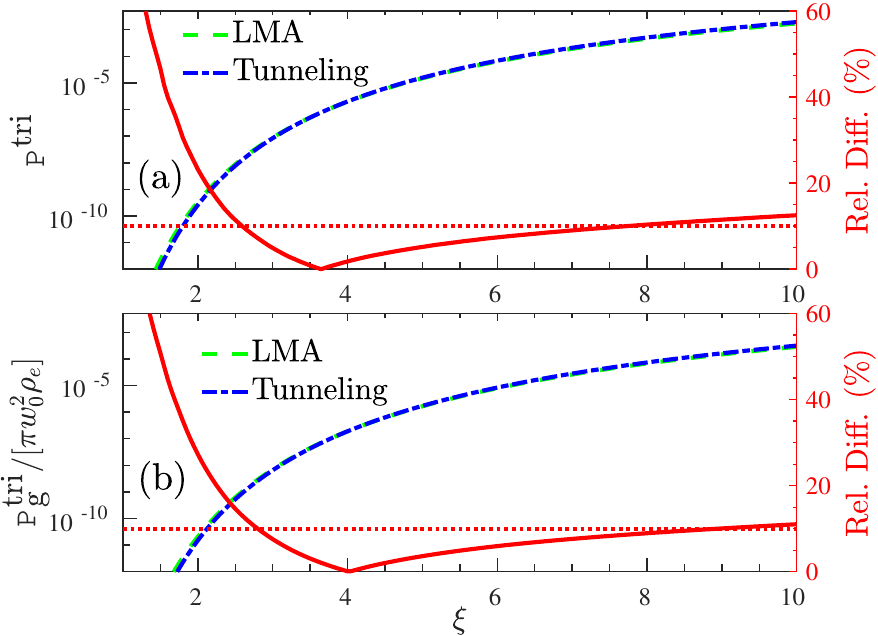}}
         \caption{Positron yield created in a plane wave (a) and a Gaussian beam (b) with $16$ cycles and intensity $\xi$ calculated by the LMA and tunnelling results. The relative difference (red solid line) between the two methods: $|\tsf{P}^{\tiny\tsf{tri}}_{\tsf{lma}}-\tsf{P}^{\tiny\tsf{tri}}_{\tsf{tun}}|/\tsf{P}^{\tiny\tsf{tri}}_{\tsf{lma}}$ is shown with the right vertical axis in each panel, and the red dotted lines denote the relative difference at $10\%$.  }
\label{Fig_two_trident}
\end{figure}

In Fig.~\ref{Fig_two_trident} (a), we show the pair creation via this two-step trident process in a plane wave background triggered by an electron with the parameter $\eta_{e}=0.2$, corresponding to the energy of $16.8~\trm{GeV}$.
Again, the LMA and the tunneling rate predict the same trend of the pair creation with the increase of the laser intensity.
Similar to the cases starting with the photon beam, the numerical comparison in Fig.~\ref{Fig_two_trident} (a) suggests intensities within $2.5\lesssim \xi \lesssim 7.5$, a `transition regime', for the measurement of this non-analytic dependency with the relative difference smaller than $10\%$.
For this two-step trident process, the optimal intensity for the smallest difference becomes slightly lower than $\xi=4$.
This is because the dominant contribution for the pair creation comes from the intermediate photon with the energy $\eta  = 0.6\eta_{e}=0.12$~\cite{PRD096004}, slightly larger than that in Fig.~\ref{Fig_ICS_source_Gausss}.

In Fig.~\ref{Fig_two_trident} (b), we consider this two-step trident pair creation in the Gaussian beam [Eq.~(\ref{Eq_Gaussian_beam})] downstream of the electron beam, in which we assume that the electron transverse scale is much broader than that of the Gaussian beam and can thus write the total pair creation by integrating the plane-wave results over the impact parameter of the laser beam
\begin{align}
\tsf{P}^{\tiny\tsf{tri}}_{\tiny\tsf{g}}&= \iint\ud^{2}\bm{x}^{\LCperp} \rho_{e}(\bm{x}^{\LCperp}) \tsf{P}^{\tiny\tsf{tri}}[\xi(\mbf{x}^{\LCperp})] \nonumber\\
           &\approx \pi w^{2}_{0}\rho_{e}  {\int^{\infty}_{0}  \ud x ~\tsf{P}^{\tiny\tsf{tri}}\left[e^{-x} \xi \right]}\,,
\end{align}
where $\rho_{e}$ is number density of the electron beam.
Similar as the plane-wave results in Fig.~\ref{Fig_two_trident} (a), the trident pair creation in the Gaussian beam also suggests a similar `transition regime' of the intensity $3.0\lesssim \xi \lesssim 8.0$ as the optimal regime for the measurement of the tunneling pair creation. The trident results are particularly significant because the probability for pair creation in the tunneling regime is in general quite low, and producing pairs in the pulse, rather than separating the photon-generation and production mechanisms is the more efficient for producing pairs.

\section{Conclusion}~\label{Sec5}
One of the interesting aspects of strong-field quantum electrodynamics in laser pulses is that exact solutions to the Dirac equation can be used to show a dependency on the coupling that cannot be arrived at using standard tools of perturbation theory. There are two `non-perturbative' dependencies: i) an all-order dependence on the charge-field coupling, $\xi$ when $\xi \not \ll 1$ (a classical effect); ii) a non-analytic dependence on the fundamental coupling, $\alphaqed$ in the quasi-static regime (a quantum effect).  The relevant parameter in experiments that collide laser pulses with particle beams is the strong-field parameter, $\chi \propto \sqrt{\alphaqed}$. It was recently reported that the non-analytic dependency can be revealed also in nonlinear Compton scattering by making a suitable cut in the transverse momentum spectrum \cite{Kampfer:2020cbx,Acosta:2021iyu} (a similar dependency was also noted in \cite{Dinu:2018efz}); here we focussed on accessing the non-analytic scaling in experiments measuring the tunneling regime of nonlinear Breit-Wheeler pair-creation from a photon for which $\chi \ll 1$. Our analysis assumes that in the tunneling regime, a plane-wave approximation is still accurate to describe pair creation in focussed laser pulses. Because the tunneling exponent is sensitive to the exact field invariants of a given scenario, we should expect that our results apply to weakly-focussed laser pulses. Although such laser pulses have a lower peak intensity than in the strongly focussed case, they are useful in experiment as they provide broader focal spots and hence minimise variation due to pulse and beam jitter. To reach the corresponding tunneling regime, the intensity parameter must be large enough to be in the quasi-static regime (where the locally constant field approximation \cite{ritus85,Harvey:2014qla,DiPiazza:2017raw,Seipt:2020diz} applies), but not too large that $\chi \not \ll 1$, whilst also producing sufficient pairs to be measureable in experiment. To assess for which parameter regime measuring the non-analytic dependency is feasible, we employed the locally monochromatic approximation, which allows one to determine: a) when the process is in the quasi-static  regime; b) when the tunneling limit of the quasi-static regime has been reached. This allowed us to identify a regime that was relatively consistent throughout scenarios where bremsstrahlung and inverse Compton sources of photons collided with focussed laser pulses as well as in scenarios where pair-creation proceeded directly in the laser pulse in an electron beam-laser collision. For photon energies of $\sim O(10\,\trm{GeV})$ the regime of intensity parameter, $\xi$, should ideally be in the intermediate intensity regime $3\lesssim \xi \lesssim 8$, corresponding to $\chi \approx 0.5$. Lower photon energies may be used by raising the intensity parameter, but at the cost of reducing the number of pairs. The optimal parameter region can be accessed at laser-particle experiments such as LUXE and E320 and also in the most recent high intensity laser facilities \cite{danson19}.

\acknowledgments
The authors thank Tom Blackburn for providing useful data from the Ptarmigan rate generator. The authors also thank the organisers of the `LUXE physics and SFQED' workshop at the Weizmann Institute, where the idea for this paper was generated. ST acknowledges the support from the Natural Science Foundation of China, Grant No.12104428.
The work was carried out in part at Marine Big Data Center of Institute for Advanced Ocean Study of Ocean University of China.

\appendix

\section{Field invariants in nonlinear Breit-Wheeler} \label{app:finvariants}
Consider the collision of a weak EM probe, $f_{p}$ with an intense background, $F_{\tiny\tsf{bg}}$. Then the total field is:
\[
F = F_{\tiny\tsf{bg}} + f_{p}.
\]
In studies of the Breit-Wheeler process, $F_{\tiny\tsf{bg}}$ is usually a plane wave, or a perturbation away from it (e.g. a weakly-focussed Gaussian beam). In studies of the Schwinger process, $F_{\tiny\tsf{bg}}$ is a constant or slowly-varying but homogeneous electric field and $f_{p}=0$ or else $f_{p}$ is a high frequency field and one writes of `assisted Schwinger'. Breit-Wheeler is then distinguished from Schwinger on the basis of $F_{\tiny\tsf{bg}}$ being a plane wave, and hence the invariants $\mathcal{S}_{\tiny\tsf{bg}} = - F_{\tiny\tsf{bg}} \cdot F_{\tiny\tsf{bg}} / 4 $ and $\mathcal{P}_{\tiny\tsf{bg}} = -F_{\tiny\tsf{bg}}\cdot\widetilde{F}_{\tiny\tsf{bg}}/4$ being zero. A fairer comparison would be to calculate the invariant of the full field. Writing the plane-wave background and probe as:
\bea
F_{\tiny\tsf{bg}} &=& m \xi_{\tiny\tsf{bg}}\left[ \vkap^{\mu}\epsilon^{\nu}-\vkap^{\nu}\epsilon^{\mu}\right] \nn \\
f_{p,j} &=& m \xi_{p}\left[ \ell^{\mu}\eps^{\nu}_{j}-\ell^{\nu}\eps^{\mu}_{j}\right]
\eea
and choosing the lightfront polarisation basis as ${\eps_{j} = \epsilon_{j} - \vkap\, \ell \cdot \epsilon_{j} / \vkap \cdot \ell}$, with $\epsilon_{1} = \epsilon$ and $\epsilon_{2}$ a spacelike vector perpendicular to $\vkap$ and $\epsilon_{1}$, we see:
\bea
\mathcal{S} &=& -\frac{1}{4} F \cdot F = F_{\tiny\tsf{qed}}^{2} \xi_{p} \chi~\delta_{j1}\nn \\
\mathcal{P} &=& -\frac{1}{4} F \cdot \widetilde{F} = F_{\tiny\tsf{qed}}^{2} \xi_{p} \chi~\delta_{j2} \nn
\eea
where $\delta_{jk}$ refers to the polarisation state of the probe, the strong-field parameter is $\chi = \xi_{\tiny\tsf{bg}} \eta$, the energy parameter is $\eta = \vkap\cdot\ell/m^{2}$, and the QED strong field scale (sometimes referred to as the `Schwinger limit') is: $F_{\tiny\tsf{qed}} = m^{2}/e$. At high background field intensities $\xi_{\tiny\tsf{bg}} \gg 1$, it is known that the Breit-Wheeler probabilities are well-characterised by a `locally-constant' rate, which depends non-trivially only on the strong-field parameter $\chi$. Hence far from being zero for Breit-Wheeler, the EM invariants are proportional to the strong-field parameter.

\section{LMA and LCFA for nonlinear Breit-Wheeler}\label{APP_LCFA_LMA}
Here we list the formulas used in the main text for nonlinear Breit-Wheeler pair-creation. (The derivation of these forms can be found in \cite{Tang:2022a}.)

The LCFA probability rate can be written as
\begin{align}
\frac{\ud\tsf{P}_{\tiny\tsf{lcfa}}}{\ud \vphi}=&\frac{\alphaqed}{\eta } \int^{1}_{0}\ud t \left[ \textrm{Ai}_1(z)- \textrm{Ai}'(z) \frac{s^{2} + (1-t)^{2}}{t(1-t)z} \right]
\label{Eq_LCFA_NBW}
\end{align}
where the argument of the Airy functions is given as $z=[s(1-s)\eta |\xi(\vphi)|]^{-2/3}$, and $t=\vkap\cdot q/\vkap\cdot \ell$ is the fraction of the lightfront momentum taken by the positron from the incoming photon, and $t\approx E_{q}/\omega_{\gamma}$ approximates as the energy ratio for nearly head-on collisions.

In linearly polarised laser backgrounds, the LMA probability rate is expressed as
\begin{align}
\frac{\ud \tsf{P}_{\tiny\tsf{lma}}}{\ud \vphi}&=\frac{\alphaqed}{\eta} \int^{1}_{0}\ud t \int^{\pi}_{-\pi}\frac{\ud \psi}{2\pi} \sum_{n=\lceil n_{\ast} \rceil}\label{Eq_LMA_NBW}\\
        &\left[\xi^{2}(\vphi)\left(\Lambda^{2}_{1,n} -\Lambda_{0,n}\Lambda_{2,n}\right)\frac{t^2+(1-t^2)}{2t(1-t)}+ \Lambda^{2}_{0,n} \right]\,,\nonumber
\end{align}
where $n_{\ast}=[1+ \xi^{2}(\vphi)/2]/[2\eta (1-t) t]$, $\Lambda_{j,n}(\zeta,\beta)$ is the generalised Bessel functions defined as
\[\Lambda_{j,n}(\zeta,\beta)=\int_{-\pi}^{\pi}\frac{\ud\phi}{2\pi}\cos^{j}(\phi)e^{i\left[n\phi-\zeta\sin(\phi)+\beta\sin(2\phi)\right]}\,,\]
with $j=0,1,2$, and the arguments \mbox{$\zeta = [\xi(\vphi) r_{b,n}\cos\psi]/[\eta t(1-t)]$}, \mbox{$\beta= \xi^{2}(\vphi) /[8\eta  t (1-t)]$}, and \mbox{$r_{b,n}  = \sqrt{2n\eta (1-t) t - 1 - \xi^{2}(\vphi)/2}$}.

\bibliography{schwingerLike}

\end{document}